\documentclass[journal]{IEEEtran}

\usepackage{epsfig,amsmath,amssymb,epsf,amsthm,scalefnt,multirow,dsfont,subfig}
\usepackage{xcolor}
\usepackage{float}
\usepackage{cite}

\newtheorem{lemma}{Lemma}

  \def\cC{{\mathcal{C}}}

 \def\cN{{\mathcal{N}}}

\def\b0{{\pmb{0}}} 

 \def\bff{{\mathbf{f}}}  \def\bh{{\mathbf{h}}}

\def\bu{{\mathbf{u}}}   \def\bx{{\mathbf{x}}}

\def\bA{{\mathbf{A}}}   
   
\def\bI{{\mathbf{I}}}   
   
 \def\bR{{\mathbf{R}}}  
\def\bU{{\mathbf{U}}}

\begin{document}

\title{\huge{Bounds on Eigenvalues of a Spatial Correlation Matrix}}

 %author names and IEEE memberships
\author{Junil Choi and David J. Love\\
\thanks{Junil Choi and David J. Love are with the School of Electrical and Computer Engineering, Purdue University, West Lafayette, IN (e-mail: choi215@purdue.edu, djlove@purdue.edu).}
}
%\markboth{Draft}%
%{Paper Outline}

\maketitle

%\squeezeup\squeezeup\squeezeup\squeezeup
\begin{abstract}
It is critical to understand the properties of spatial correlation matrices in massive multiple-input multiple-output (MIMO) systems.  We derive new bounds on the extreme eigenvalues of a spatial correlation matrix that is characterized by the exponential model in this paper.  The new upper bound on the maximum eigenvalue is tighter than the previous known bound.  Moreover, numerical studies show that our new lower bound on the maximum eigenvalue is close to the true maximum eigenvalue in most cases.  We also derive an upper bound on the minimum eigenvalue that is also tight.  These bounds can be exploited to analyze many wireless communication scenarios including uniform planar arrays, which are expected to be widely used for massive MIMO systems.
\end{abstract}

\begin{keywords}
Spatial correlation matrix,  maximum/minimum eigenvalue, exponential model, massive MIMO.
\end{keywords}
\section{Introduction}\label{sec1}
Spatial correlation in the channel can give both gains and losses depending on the scenario in multiple-input multiple-output (MIMO) systems \cite{exp_model_bruno}.  Spatial correlation is harmful for single-user MIMO systems using multiplexing because the correlation reduces the rank of the communication channel, resulting in a reduced number of parallel paths for spatial multiplexing \cite{exp_model_capa_analysis2}.  On the other hand, spatial correlation is beneficial for multi-user MIMO systems because the strong directivity of channels between a transmitter and users can help to reduce inter-user interference even with simple precoding strategies at the transmitter \cite{exp_model_bruno}.

The most common model for the spatial correlation matrix is the exponential model \cite{exp_model_bruno,exp_model_experiment}.  The exponential model is very simple because the correlation matrix is controlled by one parameter.  Although simple, it has been shown experimentally that the exponential model characterizes uniform linear array (ULA) antenna scenarios well \cite{exp_model_experiment}.  Thus, many works, e.g., capacity analyses in \cite{exp_model_capa_analysis1,exp_capa}, codebook designs for channel state information (CSI) quantization in \cite{tm_correlated5,spatial_codebook1,spatial_codebook2}, and training signal designs for channel estimation in \cite{training2,training5,cl_training_jstsp}, are based on the exponential model for the spatial correlation matrix.

The exponential model is useful for analyzing uniform planar array (UPA) scenarios.  Note that UPA deployments are growing in popularity due to the emergence of massive MIMO systems \cite{massive_mimo3,fdmimo}.  It was shown in \cite{dawei} that the spatial correlation matrix of a UPA can be approximated by the Kronecker product of the spatial correlation matrices corresponding to the vertical and horizontal domain.  In \cite{upa_codebook,dawei}, this approximation was exploited to design codebooks for CSI quantization in a UPA scenario.

Because of the reasons above, we focus on spatial correlation matrices following the exponential model in this paper.  The maximum and minimum eigenvalues of the spatial correlation matrix are important factors because they determine performance in spatially correlated channels \cite{vasan1,vasan2}.  In this paper, we derive new upper and lower bounds on the maximum eigenvalue and an upper bound on the minimum eigenvalue of the correlation matrix.  Although the exact eigenvalues of the exponential model are derived in \cite{bound_old}, the expressions need numerical solutions of the trigonometric function.  Moreover, the bounds derived from the exact expressions are not functions of the number of transmit antennas, which makes it hard to analyze massive MIMO systems with practical numbers of antennas.  Thus, it is desired to have simple and tight upper and lower bounds on extreme eigenvalues expressed with the number of antennas for analyzing the exponential model.

The new upper bound on the maximum eigenvalue, which is based on a novel matrix expansion approach, is tighter than the one from \cite{bound_old}.  Moreover, simulation results show that our lower bound on the maximum eigenvalue is very tight with the true value in general.  The new upper bound on the minimum eigenvalue is tight as well.  All these new bounds are functions of the number of transmit antennas.  The new lower bound on the maximum eigenvalue and upper bound on the minimum eigenvalue are intuitive and simple to be derived; however, we could not find such derivations even after extensive literature search.  Most of the literature adopted the previous bounds in \cite{bound_old} for performance analysis \cite{exp_model_capa_analysis1,cl_training_jstsp} or simply performed numerical studies with the exponential model \cite{exp_capa,tm_correlated5,spatial_codebook1,spatial_codebook2,training2,training5}.

\section{System Model}\label{sec2}
We consider multiple-input single-output (MISO) channels that are spatially correlated at the transmitter side. %\footnote{Our results can be easily extended to MIMO channels that are spatially correlated at both the transmitter and the receiver as well.}
%We assume that the transmitter is deployed with $N_v \times N_h$ planar antenna array as in Fig. xxxx.  The total number of antennas is given as $N_t=N_v N_h$.
Assuming $N_t$ transmit antennas at the transmitter, the input-output relation at baseband is given as
\begin{equation*}
  y=\bh^H \bx + n,
\end{equation*}
where $y$ is the received signal, $\bh\in\mathbb{C}^{N_t}$ is the channel vector, $\bx\in\mathbb{C}^{N_t}$ is the transmitted signal, and $n \sim \cC\cN(0,\sigma^2)$ is additive complex Gaussian noise.  Because we consider spatially correlated channels, $\bh$ is modeled as %$\bh = \bR^{\frac{1}{2}}\bh_w$
\begin{equation*}
  \bh = \bR^{\frac{1}{2}}\bh_w
\end{equation*}
where $\bR=E[\bh\bh^H]$ is the spatial correlation matrix and $\bh_w \sim \cC\cN(\mathbf{0},\bI_{N_t})$ is an i.i.d. complex Gaussian vector.

Depending on antenna structure, $\bR$ can be modeled in many ways.  There is particular interest of antenna deployments using UPAs in massive MIMO systems.  Note that a UPA can support a large number of antennas compactly, which makes massive MIMO practical.  For a UPA, \cite{dawei} showed by eigenvalue and capacity distributions that the spatial correlation matrix of a UPA can be approximated as
\begin{equation}\label{kron_approx}
  \bR\approx \bR_h \otimes \bR_v
\end{equation}
where $\otimes$ is the Kronecker product and $\bR_v$ and $\bR_h$ are the spatial correlation matrices of the horizontal and vertical domains, respectively.

The simplified model in \eqref{kron_approx} makes it possible to adopt any kind of ULA spatial correlation model for $\bR_h$ and $\bR_v$ because a UPA is simply a ULA in each vertical and horizontal domain (assuming co-polarized antennas).  Therefore, it is important to understand the spatial correlation matrix properties of a ULA to understand those of a UPA.

Note that the maximum and minimum eigenvalues of $\bR$ are particularly important in analyzing spatially correlated channels \cite{vasan1,vasan2}.  From \eqref{kron_approx}, we have
\begin{align*}
\bR &\approx \bR_h \otimes \bR_v\\
& = \left(\bU_h \mathbf{\Lambda}_h \bU_h^H\right) \otimes \left(\bU_v \mathbf{\Lambda}_v \bU_v^H\right)\\
& = \left(\bU_h \otimes \bU_v\right) \left(\mathbf{\Lambda}_h \otimes \mathbf{\Lambda}_v\right) \left(\bU_h \otimes \bU_v\right)^H
\end{align*}
where $\bR_h = \bU_h \mathbf{\Lambda}_h \bU_h^H$ and $\bR_v = \bU_v \mathbf{\Lambda}_v \bU_v^H$ denote the eigen-decompositions.  Let $\lambda_k(\bA)$ be the $k$-th largest eigenvalue of the matrix $\bA$.  Then, we have
\begin{align}
\lambda_1(\bR) \approx \lambda_1(\bR_h)\lambda_1(\bR_v),\quad \lambda_{N_t}(\bR) \approx \lambda_{N_t}(\bR_h)\lambda_{N_t}(\bR_v).\label{planar_max_approx}
\end{align}
For these reasons, we focus on the spatial correlation matrix of a ULA in this paper.

We let $\bR$ denote a ULA spatial correlation matrix.  There are many ways to model $\bR$ depending on scenario.  The most common and easy way to model $\bR$ is to rely on the exponential model which is given as\footnote{The field tests from \cite{exp_model_experiment} show that the exponential model characterizes the spatial correlation of ULA very well even with its simplicity.}
\begin{align}\label{exp_model}
\bR_{[i,j]}=\begin{cases}
r^{|i-j|} & \quad \text{if } i\geq j\\
(r^*)^{|i-j|} & \quad \text{if } i< j
\end{cases}
\end{align}
where $^*$ denotes a complex conjugate, $r=ae^{j\theta}$ is the correlation coefficient of $\bR$ with $0\leq a < 1$.  Note that the eigenvalues of $\bR$ only depend on the value of $a$, and $\theta$ only controls the eigenvectors of $\bR$.  Because we are interested in the maximum and minimum eigenvalues of $\bR$, we assume $r=a$ throughout the paper.

\section{Bounds on Eigenvalues}\label{sec3}
We first briefly recall previous results on analyzing the eigenvalues of $\bR$.  We then derive new bounds on the maximum and minimum eigenvalues of $\bR$.

\subsection{Previous Results}\label{prev_bounds}
In \cite{bound_old}, it has been shown that all eigenvalues of $\bR$ can be exactly derived as
\begin{equation}\label{R_eig_exact}
\lambda_i(\bR) = \frac{1-a^2}{1+a^2+2a\cos \phi_i}
\end{equation}
where $\phi_i\neq n\pi$ for any arbitrary integer $n$ are the solutions of the trigonometric equation
\begin{equation*}
\tan N_t \phi_i = \frac{-\sin \phi_i}{\left(\frac{1+a^2}{1-a^2}\right)\cos \phi_i +\frac{2a}{1-a^2}}.
\end{equation*}
From \eqref{R_eig_exact}, it is obvious that\footnote{The bounds in \eqref{max_upper1} and \eqref{min_lower} are reciprocal.  This comes from the fact that $\lambda_1(\bR)$ and $\lambda_{N_t}(\bR)$ have an approximate inverse relation regarding $a$, i.e., $\lambda_1(\bR)$ ($\lambda_{N_t}(\bR)$) increases (decreases) as $a$ grows larger.}
\begin{align}
\lambda_{1}(\bR)&\leq \frac{1+a}{1-a},\label{max_upper1}\\
\lambda_{N_t}(\bR)&\geq \frac{1-a}{1+a}.\label{min_lower}
\end{align}

\noindent \textbf{Remark:} Note that the bounds in \eqref{max_upper1} and \eqref{min_lower} are not functions of the number of antennas $N_t$.  Thus, it is not clear how the extreme eigenvalues would behave as $N_t$ grows large, which is an important aspect in predicting performance of massive MIMO systems \cite{sam}.
\subsection{New Bounds on Eigenvalues}
In the following, we derive an improved upper bound and new lower bound on $\lambda_1(\bR)$ that are both functions of $N_t$.
\begin{lemma}\label{max_bound}
With the exponential model of $\bR$ as in \eqref{exp_model}, the maximum eigenvalue of $\bR$ is bounded as
\begin{equation*}
  \frac{1+a}{1-a}-\frac{2a(1-a^{N_t})}{N_t(1-a)^2}\leq \lambda_1(\bR) \leq \frac{(1+a)(1-a^{N_t-1})}{1-a}
\end{equation*}
when $N_t>1$.
\end{lemma}
\begin{IEEEproof}
We first prove the upper bound.  We extend $\bR$ to an $2(N_t-1)\times 2(N_t-1)$ circulant matrix as
\begin{align*}
\bR_X=
\begin{bmatrix}
1 & a & \cdots & a^{N_t-1} & a^{N_t-2} & a^{N_t-3} & \cdots & a\\
a & 1 & \cdots & a^{N_t-2} & a^{N_t-1} & a^{N_t-2} & \cdots & a^2\\
\vdots & \vdots & \cdots & & \ddots & & & \vdots\\
a & a^2 & \cdots & a^{N_t-2} & a^{N_t-3} & a^{N_t-4} & \cdots & 1\end{bmatrix}.
\end{align*}
Note that $\bR$ is contained in the first $N_t$ rows and $N_t$ columns of $\bR_X$.  Therefore,
\begin{equation*}
  \lambda_{1}(\bR) \leq \lambda_{1}(\bR_X).
\end{equation*}
Because $\bR_X$ is a circulant matrix, the eigenvectors of $\bR_X$ are the columns of the $2(N_t-1)$ point DFT matrix.  If we let $\mathbf{1}_{N}$ be the $N\times 1$ vector with all one entries, it is easy to conclude that the dominant eigenvector of $\bR_X$ is
\begin{equation*}
  \bu_1(\bR_X) = \frac{1}{\sqrt{2(N_t-1)}}\mathbf{1}_{2(N_t-1)}
\end{equation*}
that cophases all of the entries of $\bR_X$ because of the assumption that $a$ is a real positive number.  Then, we have
\begin{align*}
\lambda_{1}(\bR_X) &= \bu_1(\bR_X)^H \bR_X \bu_1(\bR_X)\\
&=\frac{1}{2(N_t-1)}\sum_{k=1}^{2(N_t-1)}\left(\sum_{\ell_1=0}^{N_t-1}a^{\ell_1}+\sum_{\ell_2=1}^{N_t-2}a^{\ell_2}\right)\\
%&=\frac{1-a^{N_t}}{1-a}+\frac{a\left(1-a^{N_t-2}\right)}{1-a}\\
%&=\frac{1+a-a^{N_t-1}(a+1)}{1-a}\\
&=\frac{(1+a)(1-a^{N_t-1})}{1-a}.
\end{align*}
Thus,
\begin{equation*}
  \lambda_{1}(\bR) \leq \lambda_{1}(\bR_X)=\frac{(1+a)(1-a^{N_t-1})}{1-a}.
\end{equation*}

To prove the lower bound, we use the definition of the maximum eigenvalue, which follows the general inequality
\begin{equation*}
\lambda_1(\bR) \geq \bff^H \bR \bff
\end{equation*}
with an arbitrary vector $\bff$ satisfying $\|\bff\|_2^2=1$.  Because the elements of $\bR$ are all positive real numbers, the all one vector $\mathbf{1}_{N_t}$ with appropriate normalization would give a good lower bound on $\lambda_1(\bR)$.  Thus, we have
\begin{align*}
\lambda_1(\bR) &\geq \frac{1}{N_t} \mathbf{1}_{N_t}^T \bR \mathbf{1}_{N_t} \\
&=\frac{1}{N_t}\left(2\sum_{\ell=0}^{N_t-1} \sum_{k=0}^{\ell}a^k-N_t\right) \\
%&= \frac{1}{N_t}\left(2 \sum_{\ell=0}^{N_t-1} \frac{1-a^{\ell+1}}{1-a} -N_t\right) \\
%&= \frac{2}{1-a}-\frac{2a}{N_t(1-a)}\left(\frac{1-a^{N_t}}{1-a}\right)-1\\
&= \frac{1+a}{1-a}-\frac{2a(1-a^{N_t})}{N_t(1-a)^2}
\end{align*}
which finishes the proof.
\end{IEEEproof}
It is obvious that the upper bound in Lemma \ref{max_bound} improves on \eqref{max_upper1} because
\begin{equation*}
  \frac{(1+a)(1-a^{N_t-1})}{1-a}\leq \frac{1+a}{1-a}
\end{equation*}
for arbitrary $0\leq a < 1$.  Moreover, the upper and lower bounds on $\lambda_1(\bR)$ both converge to \eqref{max_upper1} as $N_t\rightarrow \infty$.  This shows that all three bounds become tight when $N_t$ is large.

It is also interesting to analyze the tightness of the upper and lower bounds on $\lambda_1(\bR)$ regarding $a$.  Let $\lambda_1^{\mathrm{diff}}(\bR)$ be the difference of the two bounds in Lemma \ref{max_bound}, which is given as
\begin{equation*}
\lambda_1^{\mathrm{diff}}(\bR)=\frac{(1+a)(1-a^{N_t-1})}{1-a}-\left(\frac{1+a}{1-a}-\frac{2a(1-a^{N_t})}{N_t(1-a)^2}\right).
\end{equation*}
With some algebra, we can show that $\lambda_1^{\mathrm{diff}}(\bR)$ is monotonically increasing with $a$. Moreover, $\lambda_1^{\mathrm{diff}}(\bR)\rightarrow 0$ as $a\rightarrow 0$ and $\lambda_1^{\mathrm{diff}}(\bR)\rightarrow N_t-2$ as $a\rightarrow 1$.  Thus, the two bounds are tight when $a$ is small while the gap becomes large as $a$ increases.\footnote{Note that $\lambda_1^{\mathrm{diff}}(\bR)\rightarrow N_t-2$ is only an asymptotic gap between the two bounds when $a\rightarrow 1$.  The two bounds converge to \eqref{max_upper1} as $N_t\rightarrow \infty$ whenever $a < 1$.}  This is verified numerically in Section \ref{sec_simul}.

Now, we derive an upper bound on the minimum eigenvalue of $\bR$.  The numerical studies in Section \ref{sec_simul} show that the lower bound in \eqref{min_lower} and the new upper bound on $\lambda_{N_t}(\bR)$ are both tight in general.
\begin{lemma}\label{min_upper_bound}
With the exponential model of $\bR$ as in \eqref{exp_model}, the minimum eigenvalue of $\bR$ is upper bounded as
\begin{equation*}
\lambda_{N_t}(\bR)\leq \frac{1-a}{1+a}+\frac{2a(1-(-a)^{N_t})}{N_t(1+a)^2}
\end{equation*}
%\begin{equation*}
%\lambda_{N_t}(\bR)\leq \begin{cases}\frac{1-a}{1+a}+\frac{2a(1-a^{N_t})}{N_t(1+a)^2}\quad \text{if~}N_t\text{~is even}\\
%\frac{1-a}{1+a}+\frac{2a(1+a^{N_t})}{N_t(1+a)^2}\quad \text{if~}N_t\text{~is odd.}\end{cases}
%\end{equation*}
\end{lemma}
\begin{IEEEproof}
We only prove when $N_t$ is even.  Similar derivation can be shown when $N_t$ is odd.  Using the definition of the minimum eigenvalue, we have the general inequality
\begin{equation*}
\lambda_{N_t}(\bR) \leq \bff^H \bR \bff
\end{equation*}
for an arbitrary vector $\bff$ with $\|\bff\|_2^2=1$.  Let $\tilde{\mathbf{1}}_{N}$ be the $N \times 1$ vector defined as
\begin{equation*}
\tilde{\mathbf{1}}_{N}=\begin{bmatrix}1,-1,1,-1\cdots,(-1)^{N-2},(-1)^{N-1}\end{bmatrix}^T.
\end{equation*}
We can upper bound $\lambda_{N_t}(\bR)$ with $\tilde{\mathbf{1}}_{N_t}$ as
\begin{align*}
  &\lambda_{N_t}(\bR) \leq \frac{1}{N_t} \tilde{\mathbf{1}}_{N_t}^T \bR \tilde{\mathbf{1}}_{N_t} \\
  & = 1 - \frac{2}{N_t}\left(\sum_{k=1}^{\frac{N_t}{2}}\left(N_t-(2k-1)\right)a^{2k-1}\sum_{k=1}^{\frac{N_t}{2}-1}\left(N_t-2k\right)a^{2k}\right).
\end{align*}
We can get the desired result after solving the above series and employing basic algebra.
\end{IEEEproof}

Using these bounds, we can also derive upper and lower bounds on the condition number of $\bR$ and on the (approximated) maximum and minimum eigenvalues of UPA spatial correlation matrix by \eqref{planar_max_approx}.

\section{Numerical Studies}\label{sec_simul}
\begin{figure}[t]
  \centering
  \includegraphics[width=0.9\columnwidth]{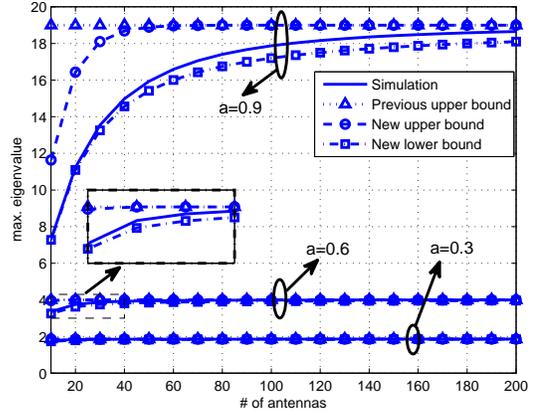}\\
  \caption{The plots of $\lambda_1(\bR)$ and its upper and lower bounds according to $N_t$.  The previous upper bound is from \eqref{max_upper1} and the new upper and lower bounds are based on Lemma \ref{max_bound}.}\label{maxeigen_fig}
\end{figure}

First, we plot the maximum eigenvalue of $\bR$, $\lambda_1(\bR)$, and the upper and lower bounds from \eqref{max_upper1} and Lemma \ref{max_bound} according to the number of antennas $N_t$ in Fig. \ref{maxeigen_fig}.  The new upper bound derived in Lemma \ref{max_bound} is tight when $a$ is low to moderate, while the gap between the new upper bound and the true $\lambda_1(\bR)$ becomes large when $a=0.9$.  However, the new upper bound keeps following the curve of $\lambda_1(\bR)$ while the previous upper bound in \eqref{max_upper1} is constant regardless of $N_t$.  It is interesting to point out that the new lower bound is tight for all values of $N_t$ and $a$.  Therefore, the new lower bound can be used as an excellent approximation of $\lambda_1(\bR)$.  As mentioned earlier, the upper and lower bounds in Lemma \ref{max_bound} are tight when $a$ is low to moderate, and the gap becomes large as $a$ approaches one.

\begin{figure}[t]
  \centering
  \includegraphics[width=0.9\columnwidth]{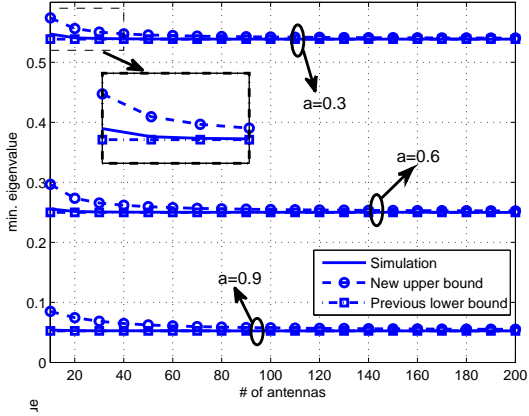}\\
  \caption{The plots of $\lambda_{N_t}(\bR)$, the previous lower bound from \eqref{min_lower} and the new upper bound from Lemma \ref{min_upper_bound} according to $N_t$.}\label{mineigen_fig}
\end{figure}

In Fig. \ref{mineigen_fig}, we plot the minimum eigenvalue $\lambda_{N_t}(\bR)$, the new upper bound from Lemma \ref{min_upper_bound}, and the previous lower bound in \eqref{min_lower} with $N_t$.  Regarding the minimum eigenvalue, the two bounds are both tight regardless of the values of $a$ and $N_t$.
\begin{figure}[t]
  \centering
  \includegraphics[width=0.9\columnwidth]{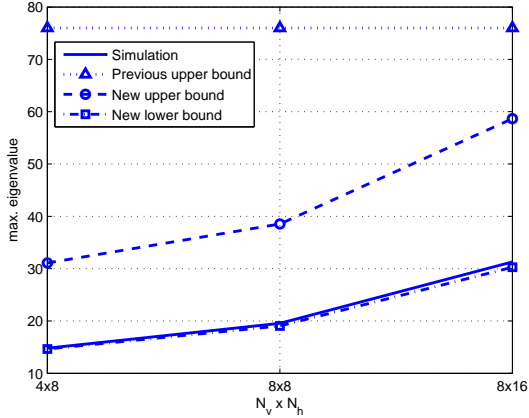}\\
  \caption{The plots of $\lambda_1(\bR)$ for the UPA scenario and its upper and lower bounds with different numbers of $N_h$ and $N_v$.  The correlation coefficients of $\bR_v$ and $\bR_h$ are set to $0.6$ and $0.9$, respectively.}\label{maxeigen_UPA_fig}
\end{figure}

Finally, we plot the maximum eigenvalue of the spatial correlation matrix of UPA given in \eqref{kron_approx} and its upper and lower bounds with different combinations of the numbers of vertical and horizontal domain antennas in Fig. \ref{maxeigen_UPA_fig}.  All bounds are based on the approximations \eqref{planar_max_approx} and derived as in the case of Fig. \ref{maxeigen_fig}.  We set the correlation coefficient of $\bR_h$ as 0.6 and that of $\bR_v$ as 0.9.  It is clear that the new lower bound is very tight for all antenna combinations.  The new upper bound also gives much better tightness compared to the previous upper bound.
%Finally, we plot the condition number of $\bR$ and two upper bounds given in \eqref{cond_bound1} and \eqref{cond_bound2} with $N_t$ in Fig. \ref{condnumber_fig}.  The plots are similar to the maximum eigenvalue case as expected.

\section{Conclusion}
In this paper, we derived new bounds on the maximum and minimum eigenvalues of a spatial correlation matrix that is characterized by the exponential model.  The upper bound on the maximum eigenvalue derived in this paper gives improved tightness than the previous upper bound.  Moreover, using numerical studies, the new lower bound on the maximum eigenvalue is shown to be very tight regardless of the number of antennas and the intensity of spatial correlation.  We also derived a new upper bound on the minimum eigenvalue of the spatial correlation matrix.  It was shown by simulations that the new upper bound and the previous lower bound on the minimum eigenvalue are both tight in general.  The theoretical results derived in this paper can be applied to performance analyses in many wireless communication scenarios including uniform planar array, which are growing in popularity due to the emergence of massive MIMO systems.

\bibliographystyle{IEEEtran}
\bibliography{refs}

\begin{thebibliography}{10}
\providecommand{\url}[1]{#1}
\csname url@samestyle\endcsname
\providecommand{\newblock}{\relax}
\providecommand{\bibinfo}[2]{#2}
\providecommand{\BIBentrySTDinterwordspacing}{\spaceskip=0pt\relax}
\providecommand{\BIBentryALTinterwordstretchfactor}{4}
\providecommand{\BIBentryALTinterwordspacing}{\spaceskip=\fontdimen2\font plus
\BIBentryALTinterwordstretchfactor\fontdimen3\font minus
  \fontdimen4\font\relax}
\providecommand{\BIBforeignlanguage}[2]{{%
\expandafter\ifx\csname l@#1\endcsname\relax
\typeout{** WARNING: IEEEtran.bst: No hyphenation pattern has been}%
\typeout{** loaded for the language `#1'. Using the pattern for}%
\typeout{** the default language instead.}%
\else
\language=\csname l@#1\endcsname
\fi
#2}}
\providecommand{\BIBdecl}{\relax}
\BIBdecl

\bibitem{exp_model_bruno}
B.~Clerckx, G.~Kim, and S.~Kim, ``Correlated fading in broadcast {MIMO}
  channels: curse or blessing?'' \emph{Proceedings of IEEE Global
  Telecommunications Conference}, Dec. 2008.

\bibitem{exp_model_capa_analysis2}
H.~Shin and J.~Lee, ``Capacity of multiple-antenna fading channels: spatial
  fading correlation, double scattering, and keyhole,'' \emph{IEEE Transactions
  on Information Theory}, vol.~49, no.~10, pp. 2636--2647, Oct. 2003.

\bibitem{exp_model_experiment}
B.~T. Maharaj, J.~W. Wallace, L.~P. Linde, and M.~A. Jensen, ``Frequency
  scaling of spatial correlation from co-located 2.4 and 5.2{GH}z wideband
  indoor {MIMO} channel measurements,'' \emph{Electronic Letters}, vol.~41,
  no.~6, pp. 336--337, Mar. 2005.

\bibitem{exp_model_capa_analysis1}
X.~Mestre, J.~R. Fonollosa, and A.~Pag\`{e}s-Zamora, ``Capacity of {MIMO}
  channels: asymptotic evaluation under correlated fading,'' \emph{IEEE Journal
  on Selected Areas in Communications}, vol.~21, no.~5, pp. 829--838, Jun.
  2003.

\bibitem{exp_capa}
S.~L. Loyka, ``Channel capacity of {MIMO} architecture using the exponential
  correlation matrix,'' \emph{IEEE Communications Letters}, vol.~5, no.~9, pp.
  369--371, Sep. 2001.

\bibitem{tm_correlated5}
J.~Choi, B.~Clerckx, N.~Lee, and G.~Kim, ``A new design of polar-cap
  differential codebook for temporally/spatially correlated {MISO} channels,''
  \emph{IEEE Transactions on Wireless Communications}, vol.~11, no.~2, pp.
  703--711, Feb. 2012.

\bibitem{spatial_codebook1}
B.~Clerckx, G.~Kim, and S.~Kim, ``{MU}-{MIMO} with channel statistics-based
  codebooks in spatially correlated channel,'' \emph{Proceedings of IEEE Global
  Telecommunications Conference}, Dec. 2008.

\bibitem{spatial_codebook2}
J.~Choi, V.~Raghavan, and D.~J. Love, ``Limited feedback design for the
  spatially correlated multi-antenna broadcast channel,'' \emph{Proceedings of
  IEEE Global Telecommunications Conference}, Dec. 2013.

\bibitem{training2}
J.~H. Kotecha and A.~M. Sayeed, ``Transmit signal design for optimal estimation
  of correlated {MIMO} channels,'' \emph{IEEE Transaction on Signal
  Processing}, vol.~52, pp. 546--557, Feb. 2004.

\bibitem{training5}
E.~Bj\"ornson and B.~Ottersten, ``A framework for training-based estimation in
  arbitrarily correlated {R}ician {MIMO} channels with {R}ician distrubance,''
  \emph{IEEE Transaction on Signal Processing}, vol.~58, no.~3, pp. 1807--1820,
  Mar. 2010.

\bibitem{cl_training_jstsp}
J.~Choi, D.~J. Love, and P.~Bidigare, ``Downlink training techniques for {FDD}
  massive {MIMO} systems: open-loop and closed-loop training with memory,''
  \emph{IEEE Journal of Selected Topics in Signal Processing}, to appear.

\bibitem{massive_mimo3}
F.~Rusek, D.~Persson, B.~K. Lau, E.~G. Larsson, T.~L. Marzetta, O.~Edfors, and
  F.~Tufvesson, ``Scaling up {MIMO}: opportunities and challenges with very
  large arrays,'' \emph{IEEE Signal Processing Magazine}, vol.~30, no.~1, pp.
  40--60, Jan. 2013.

\bibitem{fdmimo}
Y.~Nam, B.~L. Ng, K.~Sayana, Y.~Li, J.~Zhang, Y.~Kim, and J.~Lee,
  ``Full-dimension {MIMO} ({FD-MIMO}) for next generation cellular
  technology,'' \emph{IEEE Communications Magazine}, vol.~51, no.~6, pp.
  172--179, Jun. 2013.

\bibitem{dawei}
D.~Ying, F.~W. Vook, T.~A. Thomas, D.~J. Love, and A.~Ghosh, ``Kronecker
  product correlation model and limited feedback codebook design in a 3{D}
  channel model,'' \emph{Proceedings of IEEE International Conference on
  Communications}, Jun. 2014.

\bibitem{upa_codebook}
J.~Li, X.~Su, J.~Zeng, Y.~Zhao, S.~Yu, L.~Xiao, and X.~Xu, ``Codebook design
  for uniform rectangular arrays of massive antennas,'' \emph{Proceedings of
  IEEE Vehicular Technology Conference}, Jun. 2013.

\bibitem{vasan1}
V.~Raghavan, S.~V. Hanly, and V.~V. Veeravalli, ``Statistical beamforming on
  the {G}rassmann manifold for the two-user broadcast channel,'' \emph{IEEE
  Transactions on Information Theory}, vol.~59, no.~10, pp. 6464--6489, Oct.
  2013.

\bibitem{vasan2}
V.~Raghavan and V.~V. Veeravalli, ``Ensemble properties of {RVQ}-based
  limited-feedback beamforming codebooks,'' \emph{IEEE Transactions on
  Information Theory}, vol.~59, no.~12, pp. 8224--8249, Dec. 2013.

\bibitem{bound_old}
J.~N. Pierce and S.~Stein, ``Multiple diversity with nonindependent fading,''
  \emph{Proceedings of the IRE}, vol.~48, no.~1, pp. 89--104, Jan. 1960.

\bibitem{sam}
A.~Muller, A.~Kammounz, E.~Bj\"{o}rnson, and M.~Debbah, ``Efficient linear
  precoding for massive {MIMO} systems using truncated polynomial expansion,''
  \emph{IEEE Sensor Array and Multichannel Signal Processing Workshop}, to
  appear.

\end{thebibliography}

\end{document}